\magnification1200
\font\BBig=cmr10 scaled\magstep2
\font\BBBig=cmr10 scaled\magstep3


\def\title{
{\bf\BBBig
\centerline{Remarks on the solutions}
\bigskip
\centerline{of the}
\bigskip
\centerline{Maxwell- Chern-Simons theories}
}}

\def\foot#1{
\footnote{($^{\the\foo}$)}{#1}\advance\foo by 1
} 


\def\authors{
\centerline{
Z. N\'EMETH\foot{Institute for Theoretical Physics,
E\"otv\"os University, H-1088 BUDAPEST,
\hfill\break
Puskin u. 5-7 (Hungary). e-mail: nemethz@ludens.elte.hu}
}}

\def\runningauthors{N\'emeth}

\def\runningtitle{Remarks on the solutions \dots}


\voffset = 1cm 
\baselineskip = 21pt 

\headline ={
\ifnum\pageno=1\hfill
\else\ifodd\pageno\hfil\tenit\runningtitle\hfil\tenrm\folio
\else\tenrm\folio\hfil\tenit\runningauthors\hfil
\fi
\fi}

\nopagenumbers
\footline={\hfil} 


\def\and{\qquad\hbox{and}\qquad}

\def\smallcirc{{\raise 0.5pt \hbox{$\scriptstyle\circ$}}} 
\def\smallover#1/#2{\hbox{$\textstyle{#1\over#2}$}} %
\def\2{{\smallover 1/2}}

\def\parag{\hfil\break} 
\def\={\!=\!}
\def\D{{D\mkern-2mu\llap{{\raise+0.5pt\hbox{\big/}}}\mkern+2mu}\ }



\newcount\ch 
\newcount\eq 
\newcount\foo 
\newcount\ref 

\def\chapter#1{
\parag\eq = 1\advance\ch by 1{\bf\the\ch.\enskip#1}
}

\def\equation{
\eqno(\the\eq)\global\advance\eq by 1
}
\def\reference{
\parag [\number\ref]\ \advance\ref by 1
}

\ch = 0 
\eq = 1 
\foo = 1 
\ref = 1 


\title
\vskip 1.5cm
\authors
\vskip .25in

\parag
{\bf Abstract.}

{\sl The large distance behavior of the Maxwell- Chern-Simons (MCS) equations 
is analyzed, and it is found that the pure Chern-Simons limit, (when the 
Maxwell term is dropped from the equations), does not describe the large 
distance limit of the MCS model. This necessitates the solution of the original 
problem. The MCS gauge theory coupled to a nonrelativistic matter field,
(governed by the gauged non-linear Schr\"odinger equation), is studied. 
It turns out,
that there are no regular self-dual solutions as in the pure Chern-Simons case,
but the model admits interesting, though singular self-dual solutions. The 
properties of these solutions, and their large distance limits are analyzed.} .
\bigskip
\noindent
PACS numbers: 11.10.Lm, 11.10.Jj, 11.27.+d, 11.15.-q
\bigskip
%
\noindent{
(\the\day/\the\month/\the\year)}
\bigskip

\vfill\eject

It is a well known fact, that in the 2+1 dimensional electrodynamics, the 
Lagrange density of the gauge field can be modified by adding the 
Chern-Simons term
$(\kappa/4)\epsilon^{\mu\nu\alpha}F_{\mu\nu}A_\alpha$. 
This modification has many interesting
consequences, like e.g. the existence of self-dual vortex solutions
in the pure Chern-Simons limit, (when the gauge dynamics is governed
by the CS term only)[1,2,3,4,5]. These self-dual solutions have many important
applications in solid state physics, e.g. in the theory of fractional
Quantum Hall Effect [6]. The Chern-Simons modified electrodynamics may describe 
anyons, and give a method for the perturbative description of the 
Aharonov-Bohm scattering [7].
Since in a real situation we must suppose the presence
of the Maxwell term, it is an interesting task to find self-dual solutions
of the Maxwell- Chern-Simons (MCS) system. The problem becomes 
more interesting, if
we recognize, that the pure CS limit in general can not describe 
the large distance behavior of the MCS theory, as it was believed earlier.

In the first part of this paper the asymptotic properties 
of the MCS system are studied, and it is shown, 
that most of the known solutions of pure CS theory do not solve
the large distance limit of the original field equations. Then I look for 
self-dual solutions in an example of the MCS theories, in which the gauge field
is coupled with a nonrelativistic scalar field, described by the non-linear 
Schr\"odinger equation (NLSE). The same self-duality equations are used, 
which proved to be so useful in the examinations of the pure CS theories. 
It turns out, that only at very special values of the coupling constants, 
(when special relations exist between them), 
can one find self-dual solutions, but even these solutions 
are necessarily singular.
Thus the model, governed by the NLSE and the MCS system, has no such 
non singular, finite energy self-dual
solutions, as the solitons found by Jackiw and Pi in the pure CS case. 
Nevertheless there exist very interesting, though irregular solutions.

Self-duality in the nonrelativistic MCS theory has 
been already studied by G.V. Dunne and C.A. Trugenberger[8]. They introduce an 
extra neutral scalar field to find self-dual solitons in the model. In the case
of the applications in solid state physics I am interested in the existence of
self dual solutions without any extra fields. 

One can say, that the pure CS limit is 
physically the large distance and low energy limit of 
the MCS model, where the lower-derivative CS term dominates the 
higher-derivative Maxwell term. It seems plausible, looking at the power like
behavior of the vortex solutions for large values of the spatial coordinates,
(this behavior is found in many models [3,4,5])
but actually it is misleading. The pure CS theory in general can't be 
thought of as the large 
distance limit of the Maxwell-CS theory, e.g. for vortex solutions decreasing
as $r^{-n}$ at the spatial infinity, this truncation is invalidated at large 
distances. I show this for static system, because I will study static self-dual
solutions in what follows, but the results are valid in general.

The field equations for the electromagnetic field, which contain both the 
Maxwell, and the Chern-Simons terms can be written as:
$$
\matrix{\partial_iE^i-\kappa B=e\rho,
\cr
\epsilon_{ij}\partial_0 E^j+\partial_iB+\kappa E_i=-e\epsilon_{ij}J^j,
\cr}
\equation
$$
where $i,j=1,2$ , $E^i=-\partial_0 A^i-\partial_i A^0$ and 
$B=\vec{\nabla}\times\vec{A}=\epsilon_{ij}\partial_i A^j,$
and the indices are raised and lowered with the aid of the flat Minkowski 
metric with signature (+,-,-).

The static equations of the gauge field can be written as:
$$
\matrix{
\partial_i\partial^i A^0-\kappa B=e\rho,
\cr
\partial_iB+\kappa\partial_iA^0=-e\epsilon_{ij}J^j.
\cr}
\equation
$$
If we want to neglect the first terms on the left hand sides in these 
equations, to get the pure CS limit, they have to be much smaller for large 
$r$-s, than the second terms. For the soliton solutions, when every quantity is
decreasing as $r^{-n}$ in the large $r$ limit, 
these conditions can be written as:
$$
\matrix{
{\partial_i\partial^i A^0\over B}\propto{r^{-2}A^0\over B}\propto{r^{-k}},
\cr
{\partial_iB\over{\partial_iA^0}}\propto{B/r\over{A^0/r}}\propto{r^{-l}},
\cr}
\equation
$$
with k and l positive integers. From these conditions it follows, that:
$$
1={A^0\over B}\cdot {B\over A^0}\propto{r^{-(k+l-2)}}
\qquad\Rightarrow\qquad k=l=1.
\equation
$$
Thus the first terms in (2) can be neglected only if $B\propto A^0/r$
for large $r$-s. It is a very special condition, and there 
is no general reason why it should to be satisfied. (E.g. for the solutions 
in [4] and [5] this condition is not satisfied, and there is no reason to be 
satisfied for the nontopological solitons in [3].) 
In such situation
only one of the two relations in (3) may apply but not both of them. 
Therefore, in general, we can not neglect the first terms in the two 
equations of (2) at the same time. 

Now couple the gauge field, described by eq. (1), 
to a matter field, especially to a nonrelativistic 
matter field which is described by the nonlinear Schr\"odinger equation (NLSE):
$$
iD_0\psi=-{1\over2m}\vec{D}^2\psi-g|\psi|^2\psi,
\equation
$$
with
$$
D_{\mu}=\partial_{\mu}+ieA_{\mu},\qquad (\mu=0,1,2),
$$
and identify the charge density $\rho$ and current $J^j$, which appear in the 
MCS equations, with the conserved \lq\lq three current" $j=(\rho,\vec{J})$, 
where:
$$\rho=|\psi|^2, \qquad J^j=-{i\over 2m}(\psi^{\ast}D^j\psi-(D^j\psi)^{\ast}
\psi).
\equation
$$
Using the identity:
$$
\vec{D}^2\psi=(D_1\pm iD_2)(D_1\mp iD_2)\psi\pm eB\psi,
\equation
$$
the Schr\"odinger equation can be written in a new form as:
$$
i\partial_0\psi=-{1\over 2m}(D_1\pm iD_2)(D_1\mp iD_2)\psi+
(\mp {e\over 2m}B+eA_0-g|\psi|^2)\psi.
$$
This equation admits static solutions, which solve the well known, 
and generally used self-duality equation:
$$
(D_1\mp iD_2)\psi=0,
\equation
$$ 
if the algebraic equation:
$$
\mp {e\over 2m}B-eA_0-g|\psi|^2=0,
\equation
$$
is also satisfied.
With $\psi$ fields satisfying the ansatz, (8), the current density, (6), 
simplifies to:
$$
J^j=\pm {1\over2m}\epsilon^{jk}\partial_k \rho.
\equation
$$
Thus, for static, self-dual solutions, the second eq. in (2) becomes:
$$
\partial_i(A^0-{1\over\kappa}B\mp{e\over 2m\kappa}\rho)=0.
\equation
$$
This equation can also be converted into an algebraic one:
$$
A^0-{1\over\kappa}B\mp{e\over 2m\kappa}\rho=const.
\equation
$$
The well known solution of the self duality, (8), connects B with $\rho$:
$$
B=\pm{1\over 2}\vec{\nabla}^2 \ln \rho.
\equation
$$
Once eq. (8) is imposed
the three equations (9,11,12), together with the Chern-Simons modified Gauss 
law, (the first eq.in (2)), determine the behavior of the fields. Notice, 
that now we have four equations, of course, because the extra equation of 
self-duality is introduced,
but we have only three unknown functions ($A^0,B$ and $\rho$).
Furthermore two of these four equations, namely (9) and (12), are algebraic,
therefore, in the case of general coefficients in these equations, any pair of
the trhee unknown functions can be expressed in terms of the third one, (e. g.
$A^0$ and $B$ in terms of $\rho$). The remaining  two equations, (13) and the
first one in (2), constitute then 
two incompatible diferential equations for this single unknown. 
This outcome can be avoided only if the coefficients in (9) and (12)
are such that the two equations become identical. 
(This requirement establishes a connection between the coupling constants  
(here: $e,g,\kappa$ and $m$), which is called the self-duality condition.)

To make eq (9) and (12) identical,
we must choose the nonlinearity $g$ to be $g=\pm{e^2\over 2m\kappa}$, 
and tune the CS coupling constant $\kappa$ to the value
$\mp 2m$. It is a very strong self-duality condition, which not only connects 
the coupling constants, but unusually fixes the value of $\kappa$. 
In the presence of the Maxwell terms we found a stronger self-duality
condition, than for the pure CS theory, because in this case eq. (11)
containes also the $B$ field in addition to $A^0$. With this
condition, it can be seen, that $g=-({e\over 2m})^2$ and the constant, which 
couples the magnetic field to the number density, is: $\pm {e\over 2m}$, 
the same as the magnetic moment of a spin one half electron.
The negative sign of $g$ means, that the self-interaction of the matter 
field is repulsive. This sign has an important effect on the solutions, as
we will see it.

Now the self-duality (13), the algebraic
eq. (9) and the CS modified Gauss-law determine the system. 
Using the algebraic relation (9), the $\rho$ and B fields may be 
eliminated from the CS modified Gauss-law, leaving a well known, solvable
differential equation for the $A^0$ field:
$$
\bigtriangleup A^0=\kappa^2 A^0.
\equation
$$
Note, that this is consistent with the results found in topologically massive
gauge theories [9], because this equation
may describe the time component of a massive
photon field. The general solution of eq (14) can be written as:
$$
A^0=\sum_{n=0}^{\infty}\exp(in\phi) (c_1^n K_n(\kappa r)+c_2^n I_n(\kappa r)),
\equation
$$
where $I_n$ and $K_n$ are the well known Bessel functions, having blowing up
singularities at $r=0$ (respectively $r=\infty$). Thus, for $c_1^n\ne 0$,
$c_2^n\ne 0$ the solutions are necessarily singular either at the origin or at
infinity.

Now $A^0$ is a known function and using eq. (9), the $B$ field may be 
eliminated from the CS modified Gauss law, 
to get the following differential equation for $\rho$:
$$
\bigtriangleup\ln\rho={e^2\over m}\rho+4mA^0.
\equation
$$
It is an \lq\lq inhomogeneous Liouville equation" with a known source $A^0$.
If we are interested in axially symmetric, vortex like solutions, we
must choose $c_{1,2}^n$ to be zero if $n>0$, and $c_2^0=0$ guarantees the
regularity of the solution at the spatial infinity. Now the $A^0$ field
is just like a two dimensional Yukawa type potential, 
(the potential of a charged vortex interacting with a massive photon field),
with singularity at the origin and decreasing 
as ${1\over r^{1/2}}\exp(-\kappa r)$ for large r. With our choice of $A^0$,
for spatial coordinates large enough, eq. (16) simplifies to the Liouville eq.
which is known to be completely solvable. 
In the resulting Liouville equation,
the relative sign of the right and left hand sides is positive, which means,
that only singular solutions exist.

(There are some arguments from numerical calculations, which suggest, that the
solutions of the complete inhomogenous eq. (16) are always singular too.)

In this paper it was shown, that the pure CS limit is not the large distance 
limit of the Maxwell-CS theories. Then the MCS-NLSE model
was studied, using the self-duality condition, (8), which is generally used 
in CS theories. Then coupled non-linear system of differential equations 
reduced to one non-linear differential equation for the number density, 
namely to the 
inhomogeneous Liouville equation. It turns out, that the self-dual 
solutions of the MCS-NLSE model, (with self-duality (8)), must be singular.
This is a surprising result, because the well known solutions 
of the pure CS theories
are self-dual ones, but in that case there were regular solutions too: 
self-duality didn't impose such a strong constraint on the couplings 
there, as in the MCS case. Here the self-duality fixes the value of the CS
coupling $\kappa$. An interesting result of this fixing is, that the constant,
which couples the magnetic field to the number density, is 
the same as the magnetic moment of a spin one half electron. We have a lot of
information about the solutions: the form of the $A^0$ field is exactly known,
and, (in the axially symmetric case), 
it corresponds to the Yukawa type potential
of a \lq\lq point like" particle. 
The $r\rightarrow \infty$ asymptotic solution, resulting from it,  
is purely magnetic. The large distance behavior of the
number density is determined by a Liouville equation, and so the solutions in
this limit are completely known.

Finally, in the light of the findings of this paper, it would be interesting
to inquire the existence of {\sl non self-dual} nonsingular solutions in the 
MCS-NLSE model.

\vskip 1cm

I would like to thank L. Palla for many helpful discussions.

\vfill\eject


\centerline{\bf\BBig References}

\bigskip

\reference
S.~K.~Paul, A.~Khare,
Phys. Lett. B {\bf 174}, 420 (1986).

\reference
I.~Hong, Y.~Kim, Y.~Pac,
Phys. Rew. Lett. {\bf 64} 2230 (1990);
R.~Jackiw, E.~Weinberg,
Phys. Rew. Lett. {\bf 64} 2234 (1990).

\reference
R.~Jackiw, K.~Lee, E.~Weinberg,
Phys. Rew. D {\bf 42} 3488 (1990).

\reference
R.~Jackiw and S.-Y.~Pi,
Phys. Rev. Lett. {\bf 64}, 2969 (1990);
Phys. Rev. D {\bf 42}, 3500 (1990);
see Prog. Theor. Phys. Suppl. {\bf 107}, 1 (1992) for a review.

\reference
C.~Duval, P.~A.~Horv\'athy and L.~Palla, 
Phys. Rev. D {\bf 52}, 4700 (1995);
Ann. of Phys. {\bf 249}, 256 (1996).

\reference
S. M. Girvin, {\it The quantum Hall Effect}, 
ed. R. E. Prange and S. M. Girvin. Chap. 10.
(Springer Verlag, New York (1986);
S. M. Girvin and A. H. MacDonald,
Phys. Rev. Lett. {\bf 58}, 1252 (1987);
S. C. Zhang, T. H. Hansson and S. Kivelson,
Phys. Rev. Lett. {\bf 62}, 82 (1989).

\reference
O.~Bergman, G.~Lozano,
Ann. of Phys. {\bf 229}, 416, (1994).

\reference
G.~V.~Dunne, C.~A.~Trugenberger,
Phys. Rev. D {\bf 43}, 1323, (1991).

\reference
S.~Deser, R.~Jackiw, S.~Templeton,
Ann. of Phys. {\bf 140}, 372, (1982).

\bye